\documentclass{ifacconf}
\def\PaperVersion{Arxiv}
\newif\ifarxiv
\newif\iffinal
\def\ArxivString{Arxiv}
\ifx\PaperVersion\ArxivString
  \arxivtrue
  \finalfalse
\else
  \arxivfalse
  \finaltrue
\fi

\usepackage{cite}
\usepackage{amsmath,amssymb}

\usepackage{graphicx,xcolor}      
\usepackage{natbib}        
\usepackage{tcolorbox}
\usepackage{url}
\usepackage{subcaption}

\newcommand{\norm}[1]{\ensuremath{\lVert#1\rVert}}

\begin{document}
\begin{frontmatter}

\title{Characterizing all locally exponentially stabilizing controllers as a linear feedback plus learnable nonlinear Youla dynamics}

\author{Luca Furieri}

\address{Department of Engineering Science,\\ University of Oxford (e-mail: luca.furieri@eng.ox.ac.uk).}

\begin{abstract}
We derive a state-space characterization of all dynamic state-feedback controllers that make an equilibrium of a nonlinear input-affine continuous-time system locally exponentially stable. Specifically, any controller obtained as the sum of a linear state-feedback $u=Kx$, with $K$ stabilizing the linearized system, and the output of internal locally exponentially stable controller dynamics is itself locally exponentially stabilizing. Conversely, every dynamic state-feedback controller that locally exponentially stabilizes the equilibrium admits such a decomposition. The result can be viewed as a state-space nonlinear Youla-type parametrization specialized to local, rather than global, and exponential, rather than asymptotic, closed-loop stability. The residual locally exponentially stable controller dynamics can be implemented with stable recurrent neural networks and trained as neural ODEs to achieve high closed-loop performance in nonlinear control tasks. 
\end{abstract}

\end{frontmatter}

\section{Introduction} 
Many off-the-shelf reinforcement learning (RL) pipelines do not inherently certify closed-loop stability. The set of all state-to-action policies contains destabilizing controllers, which can lead to instability during training.  While this may be acceptable in episodic tasks, it is problematic in the optimal control of physical dynamical systems, where a single unstable execution can damage hardware or violate safety constraints. Recent demonstrations of deep-RL controllers for agile flight and high-speed drone racing highlight the performance that learning-based feedback laws can achieve, but also raise the question of how to maintain stability during both training and deployment. 

A strategy in control and robotics is to ensure stability and safety at the optimization-algorithm level. Safe model-based RL methods start from a safe base controller and gradually expand the explored region using statistical models and Lyapunov or barrier certificates, often with probabilistic guarantees \citep{berkenkamp2017safe,dawson2023safe,newton2022stability}. Model predictive control (MPC) achieves stability and constraint satisfaction by solving, at each step, a finite-horizon optimal control problem with appropriately designed stage costs, constraints, and terminal ingredients \citep{allgower2004nonlinear,faulwasser2018economic}. Robust, adaptive, and learning-based MPC schemes incorporate online model updates or data-driven model corrections while retaining stability and constraint satisfaction  \citep{Aswani2013provably,Hewing2020learning}. These approaches have two main advantages: they exploit model information explicitly and can enforce hard constraints. A limitation is that stability and safety are often tied to specific instances of optimal control problems, which makes it challenging to optimize arbitrary task-oriented performance objectives directly.

A complementary viewpoint is to encode stability at the policy parametrization level. In robotics, residual RL methods, which learn a correction around a pre-existing baseline policy, often outperform direct policy search, because optimization is restricted to policies that inherit favorable structure from the baseline controller \citep{johannink2019residual}. The control literature provides formal counterparts of this idea. For linear time-invariant systems, the Youla–Kućera parametrization represents all internally stabilizing controllers through a single stable transfer-function parameter \citep{youla2003modern}, enabling convex synthesis of optimal linear feedback controllers \citep{anderson2019system,furieri2019input}. For nonlinear systems, extensions based on input–output operators, coprime factorizations and stable kernel representations characterize stabilizing controllers in terms of ``Youla operators'', e.g. \cite{desoer1982global,ImuraYoshikawa1997,fujimoto1998state,fujimoto1998youla,FujimotoSugie2000}. These results typically address local and global stability in the asymptotic, input-to-state or finite-gain sense, sometimes relying on operator inverses or kernel representations that are difficult to deploy in numerical policy design and RL.

There has been a recent surge of interest in revisiting such parametrizations to enable learning over stabilizing policies. For classes of discrete-time nonlinear systems, parametrizations based on the internal model control principle can represent all globally $\ell_p$ stabilizing controllers \citep{furieri2022neural,galimberti2025parametrizations}. For the challenging partially observed setup, all policies that yield Lipschitz and globally contracting closed-loop maps have been characterized in \citep{wang2022learning,barbara2023learning,barbara2025react}, and a Youla–Kućera parametrization has been developed in the contraction framework \citep{kawano2024youla}. These approaches capture important classes of stabilizing policies and then perform unconstrained numerical optimization over them. The focus is on global stability guarantees and require either an open-loop stable plant, or the knowledge of a globally stabilizing controller and/or globally convergent observer. Designing such global baselines for high-dimensional nonlinear robotic systems may be demanding. 

In practice, real-world robotics and control applications often start from a locally stabilizing controller designed on a linearization, whereas globally stabilizing controllers are rarely available. For mechanical systems described by input-affine models,  such as manipulators, mobile robots and underactuated vehicles, local state-feedback stabilizers are routinely obtained via pole-placement, proportional--integral--derivative (PID) control, or linear--quadratic regulator (LQR) design applied to the linearization, e.g.  \cite{MellingerKumar2011,StevensLewisJohnson2015}. This motivates investigating parametrizations of nonlinear locally stabilizing controllers around a linear stabilizing baseline policy.

\subsection*{Contributions}
 \begin{tcolorbox}[
  colback=gray!1.5,
  colframe=black!20,
  boxsep=1pt,
  left=1pt,
  right=1pt,
  top=1pt,
  bottom=1pt
]
Every controller that exponentially stabilizes an input-affine system is the sum of 1) a linear feedback  $u=Kx$ that stabilizes the linearized system and 2) the output of residual locally exponentially stable controller dynamics.
\end{tcolorbox}

This result leads to a  residual policy class in which every controller is locally exponentially stabilizing by construction, while leaving the performance objective and optimization method arbitrary. Compared to \cite{ImuraYoshikawa1997}, we (i) strengthen the result from asymptotic to exponential stability, (ii) provide a constructive Lyapunov proof, and (iii) specialize the parametrization to a residual form around a linear baseline. Compared to contraction-based Youla parametrizations, we require only a locally stabilizable linearization (rather than a globally stabilizing policy and/or globally convergent observer); accordingly, only the existence of a region of attraction, rather than global stability, can be guaranteed.

Further, we propose parametrizing the free locally exponentially stable dynamics with a class of stable recurrent neural network (RNN) policies, obtained by adapting linear recurrent units from \cite{orvieto2023resurrecting} to the continuous-time setting, which can be embedded in neural-ODE-based training pipelines in the sense of \cite{chen2018neural}. Finally, in a cart–pendulum obstacle-avoidance case study, we compare our locally exponentially stable-biased residual policy class with standard multi-layer perceptron (MLP) and long short-term memory (LSTM) parametrizations of comparable size. More sample-efficient training and improved task performance are observed, demonstrating that enforcing local exponential stability at the policy class level provides a beneficial inductive bias.

\emph{Notation:} We say that a function \(f:\mathbb{R}^m \to \mathbb{R}^n\) is of class \(C^1\) if it is differentiable everywhere and its first-order derivative \(f':\mathbb{R}^m \to \mathbb{R}^{n \times m}\) is continuous. For vector $x\in \mathbb{R}^n$ and matrix $A\in \mathbb{R}^{m\ times n}$, $\|x\|$ denotes the Euclidean norm and $\|A\|$ denotes the spectral norm. The notation \(\operatorname{MLP}(x_0,\phi)\), with \(x_0 \in \mathbb{R}^n\) and \(\phi \in \mathbb{R}^d\), denotes a multi-layer perceptron with input \(x_0\) and learnable parameters \(\phi\), whose width and number of layers are chosen to be compatible with the stated dimensions.

\section{Main Result}

Consider the input-affine nonlinear system:
\begin{align}
    &\dot{x}(t) = f(x(t)) + g(x(t))\,u(t), \label{eq:system}\\
    &x(0) = x_0, \nonumber
\end{align}
where $x(t)\in\mathbb{R}^n$ and $u(t)\in\mathbb{R}^m$ denote the state and input at time $t\in\mathbb{R}$, and where 
$f:\mathbb{R}^n\!\to\!\mathbb{R}^n$ and $g:\mathbb{R}^n\!\to\!\mathbb{R}^{n\times m}$ class $C^1$ functions. Since the functions $f(\cdot)$ and $g(\cdot)$ are time-independent, we drop dependence of the arguments on time from now on for notational convenience. We assume that the origin $x=0$ is an equilibrium of \eqref{eq:system} when the input is not present, that is, $f(0) = 0$; this is without loss of generality over nonzero equilibrium points through a standard change of variables \citep{khalil2002nonlinear}. Input-affine models of the form~\eqref{eq:system} capture a broad class of mechanical and robotic systems, including manipulators, mobile robots, and underactuated systems~\citep{khalil2002nonlinear}. 

For system \eqref{eq:system}, we say that the origin is \emph{locally exponentially stable}  if there exist positive constants $c,k$, and $\lambda$ such that 
\begin{equation}
    \label{exp_stab_def}
 \norm{x(t)}  \leq k \norm{x(0)}e^{-\lambda t},\quad \forall \norm{x(0)}< c\,.
\end{equation}
This property is highly desirable as it guarantees rapid convergence of the system to the desired set points from any initial condition that is sufficiently close to them. We accordingly run the following assumption.
\begin{assum}
\label{ass:stabilizability}
    The pair $(f,g)$ in \eqref{eq:system} is exponentially stabilizable, that is, there exists a class $C^1$ function $\kappa(\cdot):\mathbb{R}^n \mapsto \mathbb{R}^m$ such that $\dot{x} = f(x)+g(x)\kappa(x)$ is locally exponentially stable at the origin and $\kappa(0) = 0$.
\end{assum}

\begin{rem}
By Lyapunov converse theorems, the origin of a nonlinear system $\dot{x} = f(x)$ with $f \in C^{1}$ in a neighbourhood of the origin is exponentially stable if and only if the Jacobian $A = \left.\frac{\partial f}{\partial x}\right|_{x=0}$ is Hurwitz \cite[Corollary~4.3]{khalil2002nonlinear}. This implies that, if Assumption~\ref{ass:stabilizability} holds, the closed-loop vector field
\begin{equation*}
f_{\mathrm{cl}}(x) = f(x) + g(x)\kappa(x)\,,
\end{equation*}
is locally exponentially stable and therefore its Jacobian
\begin{equation*}
A_{\mathrm{cl}}
:= \left.\frac{\partial (f(x) + g(x)\kappa(x))}{\partial x}\right|_{x=0}
= \left.\frac{\partial f}{\partial x}\right|_{x=0}
   + g(0)\left.\frac{\partial \kappa}{\partial x}\right|_{x=0} ,
\end{equation*}
is Hurwitz, where we used $\kappa(0) = 0$. Since the nonlinear policy $\kappa(x)$ and the linear policy $\frac{\partial \kappa}{\partial x}|_{x=0} x$  induce the same Jacobian, Assumption~\ref{ass:stabilizability} is equivalent to the existence of a matrix $K$ such that
\begin{equation*}
A + g(0)K \quad \text{is Hurwitz},
\end{equation*}
or, equivalently, to stabilizability of the linear pair $(A,g(0))$ in the classical linear-systems sense. 
\end{rem}

Under Assumption~\ref{ass:stabilizability}, the goal of this paper is to characterize all state-feedback control policies defined by
\begin{equation}
    C :=\begin{cases}
    \label{eq:control_policy}
        \dot{x}_c = f_c (x_c,x)\,,\\
        u = h_c(x_c,x)\,, \quad x_c(0) = {x_{c}}_0\,,
    \end{cases}
\end{equation}
with $f_c(\cdot,\cdot):\mathbb{R}^{n_c} \times \mathbb{R}^n \mapsto \mathbb{R}^{n_c}$ and $h_c(\cdot,\cdot):\mathbb{R}^{n_c} \times \mathbb{R}^n \mapsto \mathbb{R}^m$ class $C^1$ functions, $f_c(0,0)=0$, $h_c(0,0)=0$, and ${x_c}_0 \in \mathbb{R}^{n_c}$, that make the origin locally exponentially stable for the closed-loop system \eqref{eq:system}-~\eqref{eq:control_policy}\footnote{Note that \eqref{eq:control_policy} is a static policy $u=\kappa(x)$ when $n_c= 0$.}. We are ready to present our main technical result.
\begin{thm}
\label{thm: suff and nec}
    Suppose that Assumption~\ref{ass:stabilizability} holds, and consider the control policy
    \begin{equation}
    Q :=\begin{cases}
    \label{eq:youla_policy}
        \dot{\hat{x}} = f(x)-s(x-\hat{x})+g(x)u\,,\\
        \dot{q} = f_q(q,\hat{x},x)\,,\\
        u = K\hat{x} +  D(x)h_q(q,\hat{x},x)\,, \\
        \qquad \hat{x}(0)=\hat{x}_0,\quad q(0)=q_0\,,
    \end{cases}
\end{equation}
with $f_q:\mathbb{R}^{n_q}\times \mathbb{R}^{n_{\hat{x}}} \times \mathbb{R}^n \mapsto  \mathbb{R}^{n_q}$ and $h_q:\mathbb{R}^{n_q}\times \mathbb{R}^{n_{\hat{x}}} \times \mathbb{R}^n \mapsto  \mathbb{R}^{m}$ class $C^1$ functions, and $D:\mathbb{R}^n \rightarrow \mathbb{R}^{m \times m}$ a class $C^1$ map with $\|D(x)\|\leq1$ for every $x \in \mathbb{R}^n$. Then, the following two statements hold.

\emph{(Sufficiency)} If the policy $Q$ in \eqref{eq:youla_policy} is chosen so that:
\begin{itemize}
    \item[\emph{i)}] the matrix $K \in \mathbb{R}^{m \times n}$ makes $\left(\frac{\partial f}{\partial x}|_{x=0}+g(0)K\right)$ a Hurwitz matrix,
    \item[\emph{ii)}] the system $\dot{\zeta} = s(\zeta)$ is locally exponentially stable at the origin with $s(\cdot)$ a class $C^1$ function,
    \item[\emph{iii)}] the system $\dot{q} = f_q(q,0,0)$ is locally exponentially stable at the origin,
    \item[\emph{iv)}] it holds that $f_q(q,y,y) = f_q(q,0,0)$ and $h_q(q,y,y)=h_q(q,0,0)$ for every $y \in \mathbb{R}^n$, and $h_q(0,0,0)=0$, 
\end{itemize}
then, the origin $(x,\hat{x},q)=(0,0,0)$ of the closed-loop system \eqref{eq:system}-\eqref{eq:youla_policy} is locally exponentially stable.

\emph{(Necessity)} Every control policy $C$ in \eqref{eq:control_policy} that makes the origin $(x_c,x)=(0,0)$ of the closed-loop system \eqref{eq:system}-\eqref{eq:control_policy} a locally exponentially stable equilibrium can be rewritten as a control policy $Q$ in~\eqref{eq:youla_policy}, with $D(x)=I_m$, satisfying conditions \emph{i)}--\emph{iv)} stated above.
\end{thm}

From a theoretical point of view, Theorem~\ref{thm: suff and nec} can be viewed as an extension of the state-space parametrization of all stabilizing controllers for input-affine nonlinear systems in \citep{ImuraYoshikawa1997} from asymptotic to exponential stability. Further, the asymptotic stability proof in \citep{ImuraYoshikawa1997} is sketched via an indirect appeal to results from \cite{Vidyasagar1980b}, and an explicit Lyapunov construction is not given. In contrast, our proof of Theorem~\ref{thm: suff and nec}, \ifarxiv reported in Appendix~A,\else given in the extended arXiv version \citep[Appendix~A]{furieri2026characterizing}, \fi  ~provides a self-contained argument for exponential stability by constructing a Lyapunov function for the closed-loop dynamics, and thus certifying local exponential stability. 

The term $D(x)$ modulates the direction of the control action while preserving its magnitude and closed-loop stability. More generally, $D(x,\eta)$ may depend on relevant external information $\eta(t) \in \mathbb{R}^{d_\eta}$ provided that $||D(x,\eta)|| \leq 1$ globally and at each time. As showcased in Section~\ref{sec:experiments}, Magnitude and Direction (MAD) parametrizations ~\citep{furieri2025mad} can improve sample-efficiency and generalization in practice.

Note that the parametrization in Theorem~\ref{thm: suff and nec} is universal within the class of dynamic $C^1$ policies, since it contains every dynamic state-feedback controller that renders the equilibrium locally exponentially stable. This implies, in particular, that any globally stabilizing controller – if any exists – would be included as a special case of controller \eqref{eq:youla_policy} satisfying conditions $\emph{i)}-\emph{iv)}$, whereas policies that make the equilibrium unstable are excluded by construction. 

\begin{rem}[Region of attraction and safety]
    ~\\While the parametrization itself guarantees the existence of a region of attraction, it does not prescribe its size. If enlarging the region of convergence or ensuring compliance with safety constraints is part of the design goal, the proposed parametrization \eqref{eq:youla_policy} is directly compatible with certificate-based methods—such as neural Lyapunov, barrier, and contraction techniques—for estimating and expanding regions of attraction (e.g., \cite{berkenkamp2017safe,dawson2023safe,newton2022stability}). In this paper, we focus on by-design guarantees that hold irrespective of the training procedure and the choice of cost function. 
\end{rem}

\section{Optimizing over locally exponentially stabilizing policies}

In this section, we outline how policies of the form \eqref{eq:youla_policy} can be optimized for continuous-time optimal control of input-affine systems. We consider the problem of minimizing the infinite-horizon cost
\begin{equation}
\label{eq:cost}
J(\theta)
= \mathbb{E}_{x_0\sim\mu}\!\left[\int_0^\infty 
\ell\big(x_\theta(t;x_0),u_\theta(t;x_0)\big)\,dt\right],
\end{equation}
where $\mu$ is a probability distribution and $\ell(\cdot,\cdot)$ is a stage cost. For each $x_0 \in \mathbb{R}^n$, the closed-loop mapping of \eqref{eq:system} under a parametrized policy  $u_\theta(t) =\mathcal{K}(x([0:t]);\theta)$ with  $\theta \in \mathbb{R}^d$ is denoted by $t \mapsto (x_\theta(t;x_0),u_\theta(t;x_0))$. We aim to compute a policy that locally exponentially stabilizes the origin, while minimizing the cost \eqref{eq:cost}.

To achieve this goal, we consider the proposed class of policies
\begin{equation}
\label{eq:param_Youla}
Q_\theta:\;
\begin{cases}
\dot {\hat{x}} = f(x) - s(x-\hat{x};\theta) + g(x)\,u, \quad \hat{x}(0)=r_{\hat{x}}(x_0;\theta)\,,\\
\dot q = f_q(q,\hat{x},x;\theta),\qquad \qquad \qquad~ \quad q(0)=r_q(x_0;\theta)\,,\\
u = K\hat{x} + h_q(q,\hat{x},x;\theta)\,,
\end{cases}
\end{equation}
where the parameter $\theta \in \mathbb{R}^d$ can be optimized over. By defining the augmented state $z(t) := (x(t),\hat{x}(t),q(t))$ we can rewrite the closed-loop system \eqref{eq:system}-\eqref{eq:param_Youla}
 as $\dot z(t) = F(z(t);\theta)$, $z(0)=r(x_0;\theta)$. For notational convenience, we write $F_\theta(z):=F(z;\theta)$. The mapping $F_\theta$ is a smooth vector field that can be implemented as a differentiable module, forming a neural ordinary differential equation in the sense of \cite{chen2018neural}. This means that both forward trajectories and gradients can be computed by numerically integrating~$\dot z = F_\theta(z)$ with any desired ODE solver and its adjoint system \citep{chen2018neural}.

In implementation, one commonly optimizes only a truncated cost up to time $T>0$:
\begin{equation}
J_T(\theta)\label{eq:truncated}
:= \mathbb{E}_{x_0\sim\mu}\!\left[\int_0^{T} 
\ell\big(x_\theta(t;x_0),u_\theta(t;x_0)\big)\,dt\right].
\end{equation}
Ensuring local exponential stability around the origin makes this approximation justified in the following sense. If the corresponding
closed-loop trajectory enters the stability region by some time $T_0>0$ and
remains there, then for any $T \ge T_0$ the truncated cost $J_T(\theta)$
approximates the infinite-horizon cost $J(\theta)$ up to an exponentially
small tail. Specifically, if for all $t \ge T$ we have $\|x(t)\|\le A e^{-\gamma t}$ and
$\|u(t)\|\le B e^{-\gamma t}$ and $\ell(x,u)\le M(\|x\|^p+\|u\|^p)$, then
$\int_T^\infty \ell(x(t),u(t))\,dt \le \frac{M(A^p+B^p)}{p\gamma}e^{-p\gamma T}$.\footnote{Polynomially growing costs are standard in robotic regulation and tracking tasks around a target configuration, where quadratic penalties on state deviations and control effort are routinely used}

Next, we suggest a finite-dimensional neural network approximation of locally exponentially stable Youla-based residual architectures \eqref{eq:param_Youla} as a corollary of Theorem~\ref{thm: suff and nec}. 


\begin{prop}\label{prop: LRU}
Let $K \in \mathbb{R}^{m \times n}$ be any matrix such that
\begin{equation*}
\left.\frac{\partial f}{\partial x}\right|_{x=0}+g(0)K\,,
\end{equation*}
is Hurwitz for the system \eqref{eq:system}. Fix $\varepsilon_\lambda>0$,
and let $\sigma^+:\mathbb{R}\to\mathbb{R}_{>0}$ be any smooth positive map,
for instance the softplus function
$\sigma^+(s)=\log(1+e^s)$. For any choice of trainable parameters
\begin{align*}
\textcolor{blue}{\theta}
&=
\bigl(
\textcolor{blue}{\mu^{\hat{x}}} \in \mathbb{R}^n,
\textcolor{blue}{\mu^q_{\mathtt{Re}}} \in \mathbb{R}^{n_q},
\textcolor{blue}{\mu^q_{\mathtt{Im}}} \in \mathbb{R}^{n_q},
\textcolor{blue}{B_q} \in \mathbb{R}^{n_q \times n}
\bigr),
\\
\textcolor{blue}{\phi}
&=
\bigl(
\textcolor{blue}{\phi_1} \in \mathbb{R}^{d_1},
\textcolor{blue}{\phi_2} \in \mathbb{R}^{d_2},
\textcolor{blue}{\nu} \in \mathbb{R}^{n_q},
\textcolor{blue}{\phi_3} \in \mathbb{R}^{d_3},
\textcolor{blue}{\phi_4} \in \mathbb{R}^{d_3}
\bigr),
\end{align*}
construct the matrices
\begin{align*}
\Lambda_{\hat{x}}
&= \operatorname{diag}(\lambda^{\hat{x}}_1,\ldots,\lambda^{\hat{x}}_n),
\\
\lambda^{\hat{x}}_i
&= -\varepsilon_\lambda
   -\sigma^+\!\left(\textcolor{blue}{\mu^{\hat{x}}_i}\right),
\qquad i=1,\ldots,n,
\\
\Lambda_q
&= \operatorname{diag}(\lambda^q_1,\ldots,\lambda^q_{n_q}),
\\
\lambda^q_i
&= -\varepsilon_\lambda
   -\sigma^+\!\left(\textcolor{blue}{\mu^q_{\mathtt{Re},i}}\right)
   + j\textcolor{blue}{\mu^q_{\mathtt{Im},i}},
\qquad i=1,\ldots,n_q,
\\
\Gamma
&= \operatorname{diag}(\gamma^q_1,\ldots,\gamma^q_{n_q}),
\\
\gamma^q_i
&= \varepsilon_\lambda
   +\sigma^+\!\left(\textcolor{blue}{\mu^q_{\mathtt{Re},i}}\right),
\qquad i=1,\ldots,n_q .
\end{align*}
Here $j$ is the imaginary unit. The trainable parameters of the architecture
are highlighted in \textcolor{blue}{\textbf{blue}}. Then, the system
\eqref{eq:system} in closed loop with the policy
\begin{tcolorbox}[
  colback=gray!1.5,
  colframe=black!20,
  boxsep=1pt,
  left=1pt,
  right=1pt,
  top=1pt,
  bottom=1pt
]
\begin{subequations}\label{LRU}
\begin{align}
\dot{\hat{x}}
&= f(x)-\Lambda_{\hat{x}}(\textcolor{blue}{\mu^{\hat{x}}})(x-\hat{x})
   +g(x)u,~~~
\hat{x}(0)=\textcolor{blue}{\hat{x}_0},
\label{LRU_a}
\\
\dot{q}
&= \Lambda_q(\textcolor{blue}{\mu^q_{\mathtt{Re}}},
              \textcolor{blue}{\mu^q_{\mathtt{Im}}})q
   +\Gamma(\textcolor{blue}{\mu^q_{\mathtt{Re}}})
    \textcolor{blue}{B_q}(x-\hat{x}),~~
q(0)=\textcolor{blue}{q_0},
\label{LRU_b}
\\
u
&= K\hat{x}+\nonumber\\
   &\hspace{-0.3cm}+\tanh \left(\operatorname{LSTM}(x,\textcolor{blue}{\phi_4})\right)\odot\operatorname{MLP}_{\mathrm{nb}}
    \left(
    \begin{bmatrix}
    \operatorname{Re}(q)\\
    x-\hat{x}
    \end{bmatrix},
    \textcolor{blue}{\phi_3}
    \right),
\label{LRU_c}
\end{align}
\end{subequations}
\end{tcolorbox}
is locally exponentially stable around the origin
$(x,\hat{x},q)=(0,0,0)$. The activation functions of the multilayer
perceptrons are $C^1$, and the biases of
$\operatorname{MLP}_{\mathrm{nb}}$ are fixed to zero.
\end{prop}

\ifarxiv
\begin{pf}
The policy \eqref{LRU_a}--\eqref{LRU_c} is a special case of
\eqref{eq:youla_policy}. Since $\sigma^+$ is smooth and positive, the
resulting controller dynamics are $C^1$ in their arguments and in the
trainable parameters. It remains to check conditions \emph{i)}--\emph{iv)}
of the sufficiency statement in Theorem~\ref{thm: suff and nec}.

Condition \emph{i)} holds by assumption, since $K$ is chosen so that the
linearization
\begin{equation*}
\left.\frac{\partial f}{\partial x}\right|_{x=0}+g(0)K
\end{equation*}
is Hurwitz. To verify condition \emph{ii)}, define the estimation error
$\zeta=x-\hat{x}$. From \eqref{LRU_a} and the plant dynamics, one obtains
\begin{equation*}
\dot{\zeta}
=
\Lambda_{\hat{x}}(\textcolor{blue}{\mu^{\hat{x}}})\zeta .
\end{equation*}
By construction, the eigenvalues of
$\Lambda_{\hat{x}}(\textcolor{blue}{\mu^{\hat{x}}})$ are
\begin{equation*}
\lambda^{\hat{x}}_i
=
-\varepsilon_\lambda
-\sigma^+\!\left(\textcolor{blue}{\mu^{\hat{x}}_i}\right),
\qquad i=1,\ldots,n.
\end{equation*}
Since $\varepsilon_\lambda>0$ and $\sigma^+(s)>0$ for all
$s\in\mathbb{R}$, we have
\begin{equation*}
\lambda^{\hat{x}}_i < -\varepsilon_\lambda <0,
\qquad i=1,\ldots,n.
\end{equation*}
Hence the error subsystem is globally exponentially stable.

Condition \emph{iii)} holds because, when $x=\hat{x}$, the internal
$q$-dynamics reduce to
\begin{equation*}
\dot q
=
\Lambda_q(\textcolor{blue}{\mu^q_{\mathtt{Re}}},
          \textcolor{blue}{\mu^q_{\mathtt{Im}}})q .
\end{equation*}
The eigenvalues of
$\Lambda_q(\textcolor{blue}{\mu^q_{\mathtt{Re}}},
\textcolor{blue}{\mu^q_{\mathtt{Im}}})$ are
\begin{equation*}
\lambda^q_i
=
-\varepsilon_\lambda
-\sigma^+\!\left(\textcolor{blue}{\mu^q_{\mathtt{Re},i}}\right)
+j\textcolor{blue}{\mu^q_{\mathtt{Im},i}},
\qquad i=1,\ldots,n_q.
\end{equation*}
Therefore
\begin{equation*}
\operatorname{Re}(\lambda^q_i)
=
-\varepsilon_\lambda
-\sigma^+\!\left(\textcolor{blue}{\mu^q_{\mathtt{Re},i}}\right)
<
-\varepsilon_\lambda <0,
\qquad i=1,\ldots,n_q,
\end{equation*}
so the $q$-subsystem is globally exponentially stable.

Finally, condition \emph{iv)} holds because, for all $y$,
\begin{equation*}
f_q(q,y,y)
=
\Lambda_q(\textcolor{blue}{\mu^q_{\mathtt{Re}}},
          \textcolor{blue}{\mu^q_{\mathtt{Im}}})q
=
f_q(q,0,0),
\end{equation*}
and
\begin{equation*}
h_q(q,y,y)
=
\operatorname{MLP}_{\mathrm{no\text{-}bias}}
\left(
\begin{bmatrix}
\operatorname{Re}(q)\\
0
\end{bmatrix},
\textcolor{blue}{\phi_3}
\right)
=
h_q(q,0,0).
\end{equation*}
Moreover, $h_q(0,0,0)=0$ because the biases of
$\operatorname{MLP}_{\mathrm{no\text{-}bias}}$ are fixed to zero.
The conclusion follows from Theorem~\ref{thm: suff and nec}.
\end{pf}
\else
\begin{pf}
The policy \eqref{LRU_a}--\eqref{LRU_c} is a special case of
\eqref{eq:youla_policy}. Since $\sigma^+$ is smooth and positive, the
resulting controller dynamics are $C^1$ in their arguments and in the
trainable parameters. Condition \emph{i)} of Theorem~\ref{thm: suff and nec}
holds by the assumed Hurwitz property of the linearized closed loop. For
condition \emph{ii)}, with $\zeta=x-\hat{x}$, the plant dynamics and
\eqref{LRU_a} give
\begin{equation*}
\dot{\zeta}=\Lambda_{\hat{x}}\zeta .
\end{equation*}
The eigenvalues of $\Lambda_{\hat{x}}$ are
$\lambda^{\hat{x}}_i=-\varepsilon_\lambda-\sigma^+(\textcolor{blue}{\mu^{\hat{x}}_i})<-
\varepsilon_\lambda$, and hence the error dynamics are globally
exponentially stable. For condition \emph{iii)}, when $x=\hat{x}$ the
$q$-dynamics reduce to $\dot q=\Lambda_q q$, and every eigenvalue satisfies
\begin{equation*}
\operatorname{Re}(\lambda^q_i)
= -\varepsilon_\lambda
  -\sigma^+\!\left(\textcolor{blue}{\mu^q_{\mathtt{Re},i}}\right)
< -\varepsilon_\lambda .
\end{equation*}
Therefore the $q$-subsystem is globally exponentially stable. Finally,
condition \emph{iv)} holds because substituting $\hat{x}=x=y$ gives
$f_q(q,y,y)=\Lambda_q q=f_q(q,0,0)$ and
\begin{equation*}
h_q(q,y,y)=
\operatorname{MLP}_{\mathrm{no\text{-}bias}}
\left(
\begin{bmatrix}
\operatorname{Re}(q)\\
0
\end{bmatrix},
\textcolor{blue}{\phi_3}
\right)=h_q(q,0,0).
\end{equation*}
Moreover, $h_q(0,0,0)=0$ because the biases of
$\operatorname{MLP}_{\mathrm{no\text{-}bias}}$ are fixed to zero. The
claim follows from Theorem~\ref{thm: suff and nec}.
\end{pf}
\fi
The policy architecture \eqref{LRU_a}--\eqref{LRU_c} is a continuous-time adaptation of the Linear Recurrent Units (LRUs) proposed in \cite{orvieto2023resurrecting}. Its main advantage is that the modes of the internal dynamics can be directly controlled, while their effect on the dynamics is normalized by the matrix
$\Gamma(\textcolor{blue}{\mu^q_{\mathtt{Re}}})$. This enables globally exponentially stable recurrent dynamics and facilitates well-conditioned gradient propagation. Moreover, \cite{orvieto2023resurrecting} shows that LRUs match deep state-space models on long-sequence benchmarks while retaining RNN-like inference efficiency. This makes them an expressive and computationally efficient architecture for parametrizing the free dynamics of our residual policy class. Alternative parameterizations include recurrent equilibrium networks \citep{revay2023recurrent}, which are complete in the class of contracting and Lipschitz systems \citep{wang2022learning}, and for which a continuous-time counterpart is available \citep{martinelli2023unconstrained}.
\section{Numerical Experiments}
\label{sec:experiments}

We consider the cart–pendulum system 
\begin{equation}
x = (p,\dot p,\theta,\dot\theta)^\top \in \mathbb{R}^4,
\qquad
u \in \mathbb{R},
\end{equation}
where $p$ is the cart position, $\theta$ is the pendulum angle, and $u$ is the force applied horizontally to the cart. Let $M$ and  $m$ be the cart and pendulum masses, $L$ the pendulum length, $b$ a friction coefficient, and $g$ the gravitational constant.  The functions $f(x)$ and $g(x)$ in \eqref{eq:system} are written as
\begin{align*}
&f(x) =
\begin{bmatrix}
\dot p \\
\dfrac{m L \sin\theta\, \dot\theta^{2} + m g \sin\theta \cos\theta - b\,\dot p}{M + m \sin^2\theta } \\
\dot\theta \\
\dfrac{(M + m) g \sin\theta - m L \cos\theta \sin\theta\, \dot\theta^{2} + b\,\dot p \cos\theta}{L(M + m \sin^2\theta )}
\end{bmatrix},
\\
&g(x) =
\begin{bmatrix}
0 &
\dfrac{1}{M + m \sin^2\theta } &
0 &
-\dfrac{\cos\theta}{L(M + m \sin^2\theta )}
\end{bmatrix}^\top\,.
\end{align*}

At the upright equilibrium with the cart still at the origin
\begin{equation}
x^\star = (0,0,0,0)^\top, \qquad u^\star = 0,
\end{equation}
the linearization yields
\begin{equation}
\label{eq:linearized}
 \left.\frac{\partial f}{\partial x}\right|_{x = 0}
=
\begin{bmatrix}
0 & 1        & 0                 & 0 \\
0 & -\dfrac{b}{M} & \dfrac{g m}{M}     & 0 \\
0 & 0        & 0                 & 1 \\
0 & \dfrac{b}{M L} & \dfrac{g (M + m)}{M L} & 0
\end{bmatrix}\,,~g(0) =
\begin{bmatrix}
0 \\
\dfrac{1}{M} \\
0 \\
-\dfrac{1}{M L}
\end{bmatrix}\,,
\end{equation}
which is a stabilizable pair.

The tip of the pendulum has Cartesian position
\begin{equation}
p_{\mathrm{tip}}(x)
=
\begin{bmatrix}
p + L \sin\theta \\
L \cos\theta
\end{bmatrix}
\in \mathbb{R}^2 .
\end{equation}
We place two circular obstacles with common radius \(R>0\) and safety margin \(d_{\mathrm{safe}}>0\) at fixed centers \(c_1, c_2 \in \mathbb{R}^2\). For each obstacle, let $d_i(x) = \lVert p_{\mathrm{tip}}(x) - c_i \rVert$. The obstacle-avoidance penalty is
\begin{equation*}
\varphi_i(d_i) = \begin{cases} 0, \qquad \qquad \qquad \qquad \qquad d_i \ge R+\varepsilon_{safe},\\ \big(R+\varepsilon_{safe}-d_i\big)^2, \qquad \quad R \le d_i < R+\varepsilon_{safe},\\ \varepsilon_{safe}^{2} \;+\; \beta\big(\exp\big(\kappa (R-d_i)\big)-1\big), \quad d_i < R\,, \end{cases}
\end{equation*}
and the overall stage cost is
\begin{equation}
\label{eq:stage_experiment}
\ell(x,u) = \gamma_1\, \|x\|^2 \;+\; \gamma_2 \sum_{i=1}^2 \varphi_i(d_i(x)) \;+\; \gamma_3\, u^2\,,
\end{equation}
with $\gamma_1=0.1$, $\gamma_2=50$, $\gamma_3=0.01$. A terminal cost $\gamma_T\|x(T)\|^2$ with $\gamma_T=10$ is added to discourage degenerate strategies in which the cart stops before reaching the goal.

\subsection{Results and discussion}

\begin{figure*}[t]
  \centering
  \begin{subfigure}[t]{0.495\textwidth}
    \centering
    \vspace{0pt}
    \includegraphics[width=\textwidth, trim=0 0 0 0, clip]{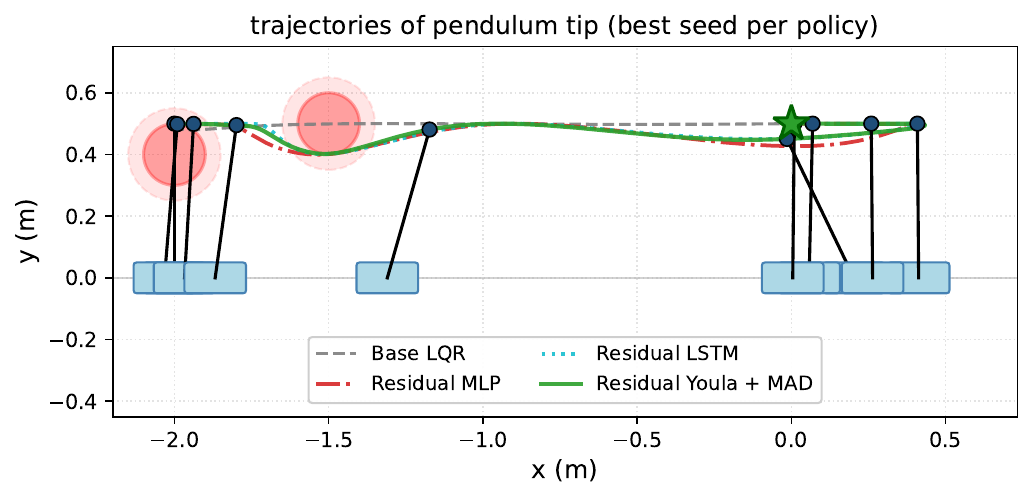}
    \caption{Trajectories of the tip of the pendulum for the best seed of each policy class. Snapshots of the cart--pendulum are shown for the proposed residual Youla + MAD policy. The green star indicates the target position. The red circles indicate obstacles, with the shadowed red area indicating the safety region around the obstacle.}
    \label{fig:tip_trajectory}
  \end{subfigure}
  \hfill
  \begin{subfigure}[t]{0.48\textwidth}
    \centering
    \vspace{0pt}
    \includegraphics[width=\textwidth, trim=0 0 0 0, clip]{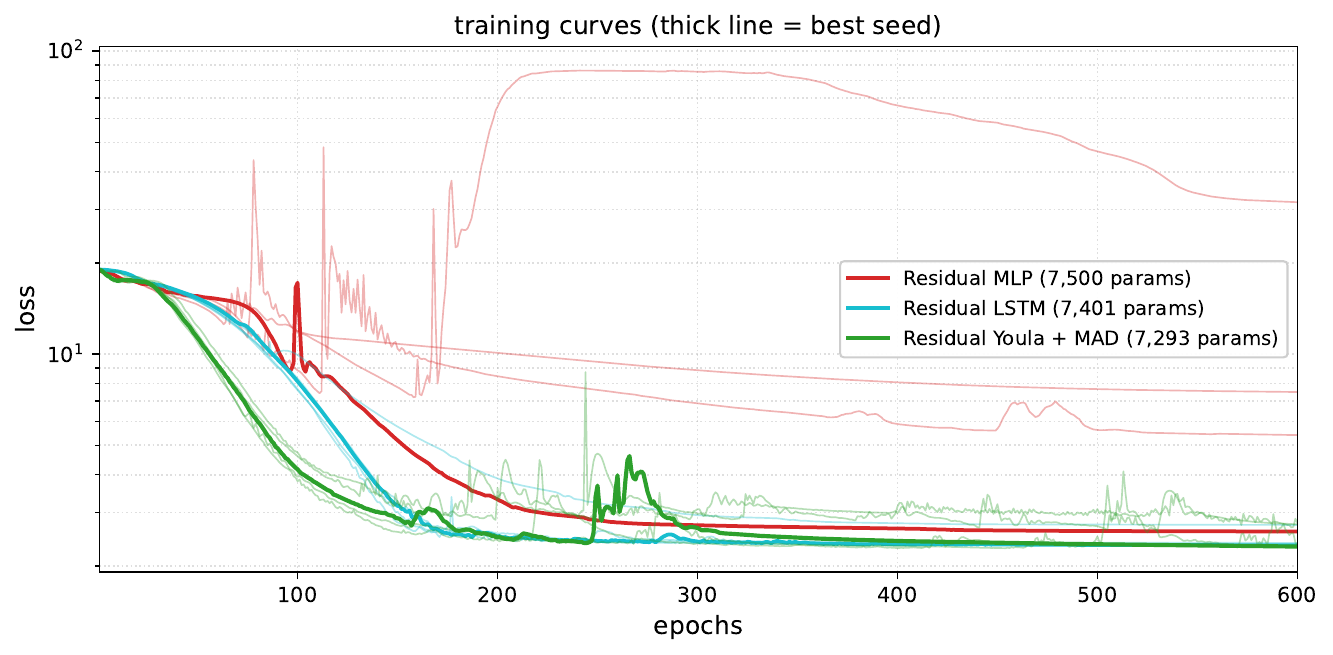}
    \caption{Training curves over $5$ seeds for each policy class. One failed seed for the MLP policy due to instability is not shown.}
    \label{fig:training_curves}
  \end{subfigure}
  \caption{Experimental results on the cart--pendulum task.}
  \label{fig:results}
\end{figure*}

First, we design a local linear stabilizer by solving the LQR problem for the linearization~\eqref{eq:linearized} with weights $Q = \mathrm{diag}(10, 1, 100, 1)$ and $R = 0.1$.

We train the proposed Residual Youla + MAD policy in \eqref{LRU_a}-\eqref{LRU_c} and two benchmark residual policies: a residual MLP $u = Kx + \operatorname{MLP}(x)$, and a residual LSTM $u = Kx + \operatorname{LSTM}(x)$. All policies have approximately $7{,}300$ trainable parameters, using Adam with identical learning rate, batch size, and epoch budget. Batches of initial conditions are drawn from a Gaussian distribution centered at $x_0 = (-2,0,0,0)^\top$ with standard deviation $0.05$. All rollouts are integrated with RK4, and policy gradients are computed by differentiating through the ODE solver~\citep{chen2018neural}.

As shown in Figure~\ref{fig:training_curves}, the proposed residual Youla + MAD policy converges to a lower training loss with fewer training episodes than both benchmarks within the same budget. Upon completing the training, the residual LSTM reaches a comparable minimal loss, but is not guaranteed to be locally exponentially stabilizing. The residual MLP fails to converge on one seed due to instability. Such seed is therefore not reported for the MLP policy. Two other seeds also incur critical cost increases.  Pendulum-tip trajectories are shown in Figure~\ref{fig:tip_trajectory}. 

These results suggest that embedding local exponential stability as an inductive bias in the policy class can improve sample efficiency and facilitate learning of complex nonlinear behaviors. The code to reproduce these results is available at \url{https://github.com/FurieriLuca/Residual-Youla}.

\begin{rem}
The goal of these experiments is to benchmark the policy class against alternatives of comparable size, rather than to develop a state-of-the-art training algorithm. In principle, all architectures considered here could be trained using more sophisticated actor-critic or model-based RL methods, improving their generalizability to wider distributions of initial conditions. In these examples, we used a basic policy-gradient training loop for all policy classes. 
\end{rem}

\section{Conclusions}

We revisited state-space nonlinear Youla-type parametrizations to characterize all dynamic state-feedback controllers that render a given equilibrium stable, extending the theory of \cite{ImuraYoshikawa1997} from asymptotic to exponential stability. The construction builds on a local linear controller that exponentially stabilizes the origin. The resulting residual policy class addresses a limitation of recent RL-oriented approaches such as \cite{furieri2022neural}, \cite{furieri2025mad}, \cite{barbara2025react}, which rely on a pre-existing globally stabilizing policy. Correspondingly, only local exponential stability can be preserved, rather than global. 

Beyond these guarantees, numerical results suggest that enforcing stability at the policy parametrization level provides a useful inductive bias for efficient training and improved generalization, compared with standard neural policies of comparable size that do not embed stability. Future directions include extending the framework to locally exponentially stabilizing output-feedback controllers for weakly detectable systems, incorporating model uncertainty and robustness requirements, and integrating Youla-based parametrizations with mechanisms for constraint handling and safety in more complex robotic systems. A broader goal is to stimulate further research on stability-aware policy representations, particularly regarding their potential for improved sample efficiency and reliable performance in challenging control tasks, clarifying when and how these parametrizations can most effectively complement RL training methods.

\bibliography{references}             

\ifarxiv
\appendix
\section{Proof of Theorem~\ref{thm: suff and nec}}    
    (\emph{Sufficiency}) Proving that the origin of the system $(x,\hat{x},q)$ is locally exponentially stable is equivalent to proving that the origin of the system $(x,e,q)$ is locally exponentially stable, where $e = x-\hat{x}$. In the coordinates $(x,e,q)$, the closed-loop interconnection of \eqref{eq:system} with a policy $Q$ in \eqref{eq:youla_policy} can be written as
    \begin{align}
        \dot{x}&=f(x)+g(x)Kx+g(x)\left[K(x-e)-Kx\right]+\label{eq:CL_x_proof}\\
        &\quad~ +g(x)D(x)\left[h_q(q,x-e,x)-h_q(q,0,0)\right]+\nonumber\\
        &\quad~ +g(x)D(x)h_q(q,0,0)\,,\nonumber\\
        \dot{e}&=s(e)\,, \label{eq:CL_e_proof}\\
        \dot{q}&=f_q(q,0,0)+\left(f_q(q,x-e,x)-f_q(q,0,0)\right)\,,\label{eq:CL_q_proof}
    \end{align}
    where we added and subtracted some terms for convenience. By denoting $\mu = (x,e,q)$ in the equation above, we equivalently write $\dot{\mu}=\overline{f}(\mu)$.

    As a special case of \cite[Th. 4.10]{khalil2002nonlinear}, local exponential stability of $(x,e,q)=(0,0,0)$ follows from finding a $C^1$ function $V:\mathbb{R}^{2n}\times \mathbb{R}^{n_q} \mapsto \mathbb{R} $ with
    \begin{align*}
    &k_1\norm{(x,e,q)}^2\leq V(x,e,q)\leq k_2 \norm{(x,e,q)}^2\,,\\ 
    &\dot{V}(x,e,q)=\nabla^\top V(x,e,q) \overline{f}(x,e,q)\leq -k_3\norm{(x,e,q)}^2\,,
    \end{align*}
    for all $(x,e,q) \in D\subset \mathbb{R}^{2n} \times  \mathbb{R}^{n_q}$, where $D$ is a domain containing the origin and $k_1$, $k_2$ and $k_3$ are positive constants. We proceed to construct such a function.

    Since all involved functions are $C^1$ and conditions $i)$, $ii)$ and $iii)$ ensure locally exponentially stability of the origin for the systems $\dot{x} = f(x)+g(x)Kx$, $\dot{e}=s(e)$ and $\dot{q}=f_q(q,0,0)$ respectively, the converse Lyapunov theorem \cite[Theorem 4.14]{khalil2002nonlinear} implies that there exist positive constants $\rho$, $c_1$, $c_2$, $c_3$ and $c_4$ and functions $V_x:B^{n}_\rho \mapsto \mathbb{R}$, $V_e:B^n_\rho \mapsto \mathbb{R} $  and $V_q:B^{n_q}_\rho \mapsto \mathbb{R} $, where $B^n_\rho = \{x\in \mathbb{R}^n|~\norm{x}\leq \rho\}$\footnote{The radii of the balls could be different, so we consider a smallest common radius $\rho>0$ for simplicity.} and $B^{n_q}_\rho = \{q \in \mathbb{R}^{n_q}|~\norm{q}\leq \rho\}$ satisfying  $\norm{h_q(q,0,0)}\leq L\norm{q}$ and
    \begin{align*}
        &c_1 \norm{z}^2 \leq V_z(z)\leq c_2 \norm{z}^2, \forall z \in \{\text{``$x$''},\text{``$e$''},\text{``$q$''}\}\nonumber\\
        &\nabla^{\top} V_x(x)(f(x)+g(x)Kx)\leq -c_3 \norm{x}^2\,,\nonumber\\
        &\nabla^{\top} V_e(e)s(e)\leq -c_3 \norm{e}^2\,,~~\nabla^{\top} V_q(q)f_q(q,0,0)\leq -c_3 \norm{q}^2\,,\nonumber\\
        &\norm{\nabla V_z(z)}\leq c_4 \norm{z},\quad \forall z \in \{\text{``$x$''},\text{``$e$''},\text{``$q$''}\}\,,\nonumber
    \end{align*}
    
     Furthermore, $\norm{g(x)}\leq G$ for all $x \in B^n_\rho$. Also, by property $\emph{iv)}$ 
 and the $C^1$ assumptions we have that, for all $(x,e) \in B^n_\rho$ there exists a constant $L$ such that
    \begin{align*}
        &\norm{K(x-e)-Kx}\leq L\norm{e}\,,\\
        &\norm{D(x)[h_q(q,x-e,x)-h_q(q,0,0)]} \leq\\ &\quad\leq\norm{D(x)}\norm{[h_q(q,x-e,x)-h_q(q,x,x)]}\leq L \norm{e}\,,\\
        &\norm{f_q(q,x-e,x)-f_q(q,0,0)}\leq L \norm{e}\,,
    \end{align*}
where we used the fact that $\|D(x)\|\leq 1$ for all $x \in \mathbb{R}^n$ and $L$ is taken as the worst-case value among the three inequalities. We proceed to compute the time derivatives of $V_x$, $V_e$ and $V_q$ over the trajectories of the closed-loop system. By exploiting the above upperbounds and the expression \eqref{eq:CL_x_proof}, we deduce that within $B^n_\rho$:
\begin{align*}
    \dot{V}_x=\nabla V^\top(x)\dot{x} \leq-c_3 \norm{x}^2+c_4\norm{x}GL(2\norm{e}+\norm{q})\,,
\end{align*}
and by applying Young's inequality $ab \leq \frac{\epsilon}{2}a^2+\frac{1}{2 \epsilon}b^2$ with $a = \norm{x}$, $b=c_4GL(2\norm{e}+\norm{q})$ and $\epsilon = c_3$, we conclude
\begin{align}
    \dot{V}_x(x)&\leq -\frac{c_3}{2} \norm{x}^2+\frac{5c_4^2 G^2 L^2}{2c_3}(\norm{e}^2+\norm{q}^2)\nonumber\\
    &=-\frac{c_3}{2} \norm{x}^2 + M(\norm{e}^2+\norm{q}^2)\,.\label{eq:bound_Vxdot}
\end{align}
By \eqref{eq:CL_e_proof} we  have that $\dot{V}_e(e) \leq -c_3 \norm{e}^2$ within $B^n_\rho$, and analogously to above, we deduce that within $B^{n_q}_\rho$:
\begin{align}
    &\dot{V}_q(q)\leq -c_3 \norm{q}^2+\nabla^\top V_q \left(f_q(q,x-e,x)-f_q(q,0,0)\right)\nonumber\\
    &\leq -\frac{c_3}{2}\norm{q}^2+c_4\norm{q}L\norm{e}\nonumber\\
    &\leq -\frac{c_3}{2}\norm{q}^2+\frac{c_4^2L^2}{2c_3}\norm{e}^2 = -\frac{c_3}{2}\norm{q}^2+M_2\norm{e}^2\label{eq:bound_Vqdot}\,.
\end{align}

Last, consider a candidate Lyapunov function  
\begin{equation}
\label{eq:total_Lyap}
    V(x,e,q) = V_x(x)+\alpha V_e(e) + \beta V_q(q)\,,
\end{equation}
with $\alpha>0$ and $\beta >0$ to be determined. By assuming that $(x,e,q) \in B^{n}_\rho \times B^{n}_\rho \times B^{n_q}_\rho$ and plugging the expressions \eqref{eq:bound_Vxdot}-\eqref{eq:bound_Vqdot} and  $\dot{V}_e(e) \leq -c_3 \norm{e}^2$ it holds that
\begin{align*}
    &\dot{V}(x,e,q) = \dot{V}_x(x)+\alpha \dot{V}_e(e) + \beta \dot{V}_q(q)\\
    &\leq \frac{-c_3}{2}\norm{x}^2 \hspace{-0.1cm}+\hspace{-0.1cm} (\beta M_2-\alpha c_3+M)\norm{e}^2\hspace{-0.1cm}+\hspace{-0.1cm}\left(M-\frac{\beta c_3}{2}\right)\norm{q}^2\,.
\end{align*}
We conclude that, by selecting $ \beta > \frac{2M}{c_3}$ and $\alpha >\frac{\beta M_2+M}{c_3}$, the function  \eqref{eq:total_Lyap} satisfies $\overline{c}_1\norm{(x,e,q)}^2 \leq  V(x,e,q)\leq \overline{c}_2\norm{(x,e,q)}^2$ and $\dot{V}(x,e,q)\leq - F \norm{(x,e,q)}^2$, where $\overline{c}_1=c_1 \min\{1,\alpha,\beta\}>0$, $\overline{c}_2=c_2\max\{1,\alpha,\beta\}>0$ and $F=\min\{\frac{c_3}{2},\frac{\beta c_3}{2}-M,\alpha c_3-\beta M_2-M\}>0$. Hence, the origin $(x,e,q) = (0,0,0)$ of the closed-loop system is locally exponentially stable over $(x,e,q) \in B^{n}_\rho \times B^{n}_\rho \times B^{n_q}_\rho$.

\smallskip

(\emph{Necessity}) Consider any policy $C$ in the form \eqref{eq:control_policy} that makes the origin $(x_c,x)=(0,0)$ of the closed-loop system a locally exponentially stable equilibrium. This means that 
\begin{align}
    &\dot{x} = f(x)+g(x)h_c(x_c,x)\,,\label{eq:CL_system_Target}\\
    &\dot{x}_c = f_c(x_c,x)\,,
\end{align}
is locally exponentially stable around the origin. Analogous to \cite{ImuraYoshikawa1997}, we first construct a policy $Q$ in the form \eqref{eq:youla_policy} that is equivalent to \eqref{eq:CL_system_Target}. Choose $D(x)$ to be the identity matrix of dimension $m$. Choose $n_q = n+n_c$ and split $q = (q_1,q_2)$ with $q_1 \in \mathbb{R}^n$ and $q_2 \in \mathbb{R}^{n_c}$. Define $\eta = q_1+x-\hat{x}$ and choose
\begin{equation*}
    f_q(q,\hat{x},x) = \begin{bmatrix}f(\eta)-s(x-\hat{x})+g(\eta)h_c(q_2,\eta)\\f_c(q_2,\eta)\end{bmatrix}\,,
\end{equation*}
and $h_q(q,\hat{x},x)=-K q_1+h_c(q_2,\eta)$. Also, select $\hat{x}(0)=q_1(0)$. Denoting $d = \hat{x}-q_1$, it holds that
\begin{align*}
    \dot{d} &= f(x)-f(x-d)+g(x)u-g(x-d)h_c(q_2,x-d)\\
    &=f(x)-f(x-d)+g(x)(K(d+q_1)-Kq_1+\\
    &\quad+h_c(q_2,x-d))-g(x-d)h_c(q_2,x-d)\,.
\end{align*}
Since $f$ and $g$ are $C^1$ functions, the differential equation above admits a unique solution $d(0) = d(t) =0$. Therefore, by choosing $\hat{x}(0)=q_1(0)$, it holds that $\hat{x}(t) =q_1(t)$  and $\eta(t) =x(t)$ at all times. By taking $q_2(0) = {x_c}(0)$ we deduce that $q_2$ and $x_c$ satisfy the same differential equation with the same initial condition, and therefore $q_2(t) = x_c(t)$ at all times. We conclude that the trajectories $u(t)=K\hat{x}-Kq_1 +h_c(q_2,\eta)$  and $u(t) = h_c(x_c,x)$ are equivalent. It remains to verify that the control policy $Q$ that we have constructed complies with the requirements $\emph{i)}$-$\emph{iv)}$. Conditions $\emph{i})$ and $\emph{ii)}$ hold by construction. For $\emph{iv})$ it holds that 
\begin{equation}
\label{eq:expr_q_dynamics}
    f_q(q,y,y) = \begin{bmatrix}f(q_1)+g(q_1)h_c(q_2,q_1)\\f_c(q_2,q_1)\end{bmatrix}=f_q(q,0,0)\,, 
\end{equation}
and $h_q(q,y,y)=-Kq_1+h_c(q_2,q_1) = h_q(q,0,0)$, and $h_q(0,0,0)=h_c(0,0) = 0$. For $\emph{iii)}$, note that the dynamics of $\dot{q} = f_q(q,0,0)$  can be expressed as per \eqref{eq:expr_q_dynamics} which coincides with the closed-loop dynamics
\begin{equation*}
    \begin{bmatrix}\dot{x}\\ \dot{x}_c\end{bmatrix} = \begin{bmatrix}f(x)+g(x)h_c(x_c,x)\\f_c(x_c,x)\end{bmatrix}\,.
\end{equation*}
Since the latter are assumed to be locally exponentially stable around the origin, so are the designed dynamics $\dot{q} = f_q(q,0,0)$.


\else
\fi

\end{document}